# Non-radiating and radiating configurations driven by left-handed metamaterials

**A. D. Boardman and K. Marinov**

Photonics and Nonlinear Science Group, Joule Laboratory, Department of Physics,

University of Salford, Salford M5 4WT, UK

Abstract

It is shown that a pair of identical emitters (e.g. wire dipole antennas) in the focal points of a disc, made of left-handed metamaterial (a "perfect" lens), form a non-radiating electromagnetic configuration. The emitters are fed with voltages of equal magnitude and $\pi$-out-of-phase. Detailed finite-difference time-domain (FDTD) modeling shows that there are non-propagating electromagnetic fields generated – fields that remain confined within the region between the emitters and the lens. The energy balance of the system shows that the radiation resistance of the system is very low. This means that the input power is converted to heat in the volume of the lens and only a small fraction of it is radiated. The system performance shows that disturbing the configuration of the non-propagating electromagnetic fields with the presence of an externally introduced object stimulates radiation. This suggests possible detector applications. In-phase feeding voltages are also studied with the consequence that the radiation resistance of the antennae is increased.

# 1. Introduction

Non-radiating electromagnetic configurations are oscillating charge-current distributions that do not produce electromagnetic radiation – i.e. their electromagnetic fields are zero in the radiation zone. Considering currents driven by voltages, as is the case with a normal wire antenna, a "non-radiating" would be a system with zero *radiation* resistance. Indeed, trivial cases, such as a wire-loop in a quasi-static regime, or a standard capacitor, are known not to radiate, and are of no interest. Important electromagnetic non-radiating systems first appeared in the context of atomic and particle physics problems [1-3]. The general conditions under which a charge-current configuration does not emit radiation have been derived in [1, 4] and the existence of nontrivial non-radiating structures has been established [1, 2]. As defined [5] a "non-radiating" source – to a partial differential equation – is a source that generates zero "field" outside its own volume. Apart from electrodynamics, the problem is of interest in other branches of physics, such as acoustics and gravitational theory [5-7]. The motivation for studying non-radiating systems is, in part, driven by the practical importance of the so-called inverse-source problems [8-11] that involve reconstructing the sources from the fields they generate. In the general case, however, the absence of additional constraints would make it impossible to reconstruct the mathematical form of a wave source from information about the fields outside the volume of this source [8]. This fundamental nonuniqueness of the solution of the inverse-source problem has been shown to be related to the existence of nontrivial "nonradiating" solutions of the corresponding source-field problem. On the other hand, a class of "purely radiating" sources, complementary to the nonradiating ones has also been identified and the possibility for decomposing the source into a sum of nonradiating and "purely radiating" parts has been investigated [7, 10]. In a way it is only the "purely radiating" component of the source that can communicate with the outside world, but, on the other hand, the nonradiating part plays a role in the internal dynamics of the system.

Non-radiating electromagnetic systems can be constructed by combining infinitesimal toroidal and supetoroidal currents with electric or magnetic dipoles [12, 13]. In particular, the electromagnetic fields of a toroidal solenoid and an electric dipole placed in the center of the toroid will compensate each other under certain conditions. It is interesting that electromagnetic potentials survive in the latter case. In fact the conditions for absence of electromagnetic potentials and electromagnetic fields are not identical [11]. The condition for the absence of radiation [1, 2] is a condition for absence of electromagnetic potentials. While the absence of electromagnetic potentials automatically ensures the absence of electromagnetic fields, the opposite is not true: in principle finite electromagnetic potentials can exist in the absence of electromagnetic fields, which means that the condition for absence of radiation [1] is not a necessary one. A necessary and sufficient condition for absence of radiation has been derived [4]. In this context it has been shown [14] that the combination of a toroidal solenoid and an electric dipole satisfies the latter necessary and sufficient condition and does *not* satisfy the more restrictive earlier condition. Moreover, any system with this property will generate non-zero electromagnetic potentials in the absence of electromagnetic fields.

The important question concerning non-radiating electromagnetic structures is what *practical* applications they might have and several possibilities have already been discussed in the literature. For example, it has been suggested that it should be possible to generate time-dependent electromagnetic potentials in the absence of electromagnetic fields [12, 13]. In general, a suitably chosen non-radiating system may be used to compensate, partially, the reactive power associated with an antenna [10], without affecting its far-field radiation pattern in any way, thus improving its performance. It has been shown recently [14, 15] that non-radiating systems in the form of a suitable combination of toroidal and supertoroidal solenoids and electric/magnetic dipoles can be used to measure the relative dielectric permittivity of the ambient material and that, the use of systems involving higher-order supertoroids may be advantageous. From a practical point of view, however, these systems suffer from certain limitations. These are connected with the fact that the radiation pattern of a toroidal or supertoroidal solenoid matches that of an electric or magnetic dipole only when the dimensions of the toroidal solenoid are much smaller than the radiation wavelength. This means that high resolution must be used is such a system is to be modeled numerically [14]. In addition the radiation resistance of an emitter which is small on a wavelength scale is low. This complicates possible experimental investigations. Besides, utilizing the "non-propagating" field, which by definition exists "inside" any non-radiating configuration, would be difficult for a system much smaller than the wavelength.

The aim of the present investigation is to show that the remarkable properties of left-handed *metamaterials* (LHM) [16-24] can be used to produce a non-radiating electromagnetic system capable of creating a non-propagating electromagnetic field on a large scale, compared to the wavelength. The analysis of this problem is extremely well-suited to the FDTD method [25] which allows the role of LHM in non-radiating configurations to be analyzed in full numerical and to emphasize features that seem to be generic for all non-radiating electromagnetic systems. The results reported later on show that a non-radiating electromagnetic system can be used as a detector because any object disturbing the non-propagating field inside the system causes the system to radiate, thus revealing its presence. It is shown that the sensitivity of such a device is quite high.

As stated earlier a non-radiating electromagnetic system is, in fact, a system with zero (in the ideal case) radiation resistance. The same principle behind the operation of such a non-radiating configuration can be exploited, permitting an increase in the radiation resistance of the common wire-dipole antenna to be achieved. This improves its performance. This increase occurs in a regime where the antenna length is considerably smaller than one-half of the wavelength (the standard half-wavelength dipole) where the antenna radiation resistance is low. Demands for miniaturization require antenna dimensions to be much lower than the operation wavelength and should low-loss LHMs become available in the future the system considered here may prove to be attractive for device construction.

## 2. System design and modeling

The system under investigation and its principle of operation are shown on Fig. 1 (a) and (b), respectively. It consists of a pair of identical wire-dipoles $d_1$ and $d_2$, each of length $L_d$, center-fed by voltages $U_1(t) = U_0 \sin(\omega t)$ and $U_2(t) = U_0 \sin(\omega t + \varphi)$, coaxial with a disc of LHM of thickness $L$ and radius $R_L$ (a "perfect" lens [16-23]). The centers of the dipoles are at the focal points of the lens and are on the system axis. At a later stage, a dielectric object with a square cross-section of size $D$ in the $\rho$-$z$ plane is introduced to the lensing arrangement, as shown in Fig. 1 (a). The relative dielectric permittivity of this object is $\varepsilon_P$, its internal radius is $R_P$ and it is located at a height $Z_P$ above the lens. Naturally, due to the cylindrical symmetry the actual shape of this object is a ring.

To create a non-radiating electromagnetic system, the voltages, feeding the antennae must be equal in magnitude and $\pi$-out-of phase ($\varphi = \pi$), which means that their individual fields will be equal in magnitude but $\pi$-out-of phase. In order to illustrate the system wire-dipoles are used. This is the simplest arrangement to analyze, but any two identical emitters can be used with the same outcomes. The only requirement is that the radiation patterns of the emitters must be symmetric in the sense that the radiation pattern of emitter 1 (Fig. 1(b)) must be symmetric with respect to the plane $O_1O_2$ and similarly for emitter 2. Identical emitters mean emitters with identical radiation patterns [11] but not necessarily *physically* identical. For example, emitter 1 could be a toroidal solenoid and emitter 2 could be a suitably chosen infinitesimal electric dipole [12-15].

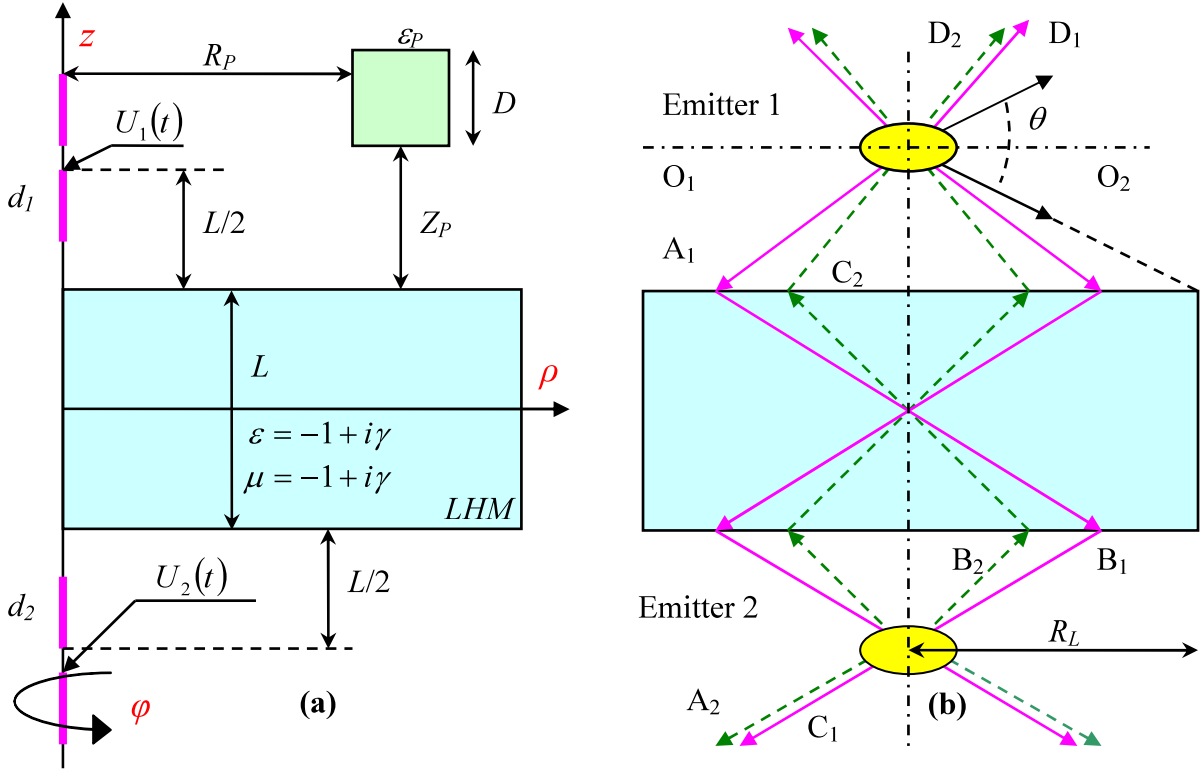

Fig. 1. (Color online) System design (a) and principle of operation (b) of a non-radiating configuration exploiting negative refraction in LHM. $d_1$, $d_2$ wire-dipoles of length $L_d$ each, center-fed by voltages $U_1(t) = U_0 \sin(\omega t)$ and $U_2(t) = U_0 \sin(\omega t + \varphi)$, coaxial with a disc of LHM of thickness $L$ and radius $R_L$ (a "perfect" lens). The centers of the dipoles are in the focal points of the lens. A square-shape (in the $\rho$-$z$ plane) perturbation made of a material with dielectric permittivity $\varepsilon_P$ of size $D$ at a distance $Z_P$ and $R_P$ from the surface of the lens and the system axis, respectively, is introduced to study its effect on the system. The lens will prevent the radiation of any two emitters, with symmetric (with respect to the plane $O_1O_2$, in the case of Emitter 1) radiation patterns, driven by $\pi$-out-of phase voltages, to escape the system (see the text).

Using the laws of *negative refraction* a ray originating at emitter 1 and following the path (emitter 1)-$A_1B_1C_1$ (Fig. 1(b)) will interfere destructively with a ray emitted by emitter 2, such as $A_2$. This is because any phase shift resulting from propagation in air, by traveling along the paths (emitter 1)-$A_1$ and $B_1C_1$, is compensated by the phase-shift resulting from the propagation along the path $A_1B_1$ in the LHM. Thus when a ray emitted by emitter 1 arrives at emitter 2 it is π-out-of phase with the corresponding ray emitted by emitter 2: $A_2$ in this case. The same reasoning is valid for any ray emitted by emitter 2 and that impacts upon the lens. For example, the ray (emitter 2)-$B_2C_2D_2$ interferes *destructively* with the ray (emitter1)-$D_1$. It is clear then, that no ray can escape the system and contribute to the radiation field, except for the rays in the angular interval $\theta$, (Fig. 1(b)) where

$$\theta = 2\tan^{-1} \frac{L}{2R_L}. \qquad (1)$$

This "escape angle" can be made arbitrarily small by increasing the radius of the lens. Alternatively, emitters with a zero in their radiation pattern along a direction perpendicular to system axis (the plane $O_1O_2$) can be used. In fact vertical electric dipoles are used here, and their radiation pattern has a *maximum* in a direction perpendicular to the system axis. Nevertheless, the system performance is very good at reasonable values of the lens radius $R_L$, as will be shown later.

The lens creates an "image" of emitter 1 superimposed on emitter 2 and vice-versa. The result of the interaction between the emitters and their images is that the electromagnetic energy remains confined in the region between the emitters and the lens and no radiation is generated. This outcome, however, does not contradict the energy conservation law because the energy radiated by the emitters is dissipated by the losses in the lens, as will be shown.

If an object with a dielectric permittivity $\varepsilon_P$ is present (such as the ring shown on Fig. 1(a)) the situation will change. Suppose that the object is on the ray-path (emitter 1)-$A_1$. Even neglecting the possible reflection and the attenuation accompanying the transmission through the object, the ray (emitter 1)-$A_1B_1C_1$ will acquire an additional phase shift, resulting from the propagation through the object. The interference of the original ray with the ray (emitter 2) - $A_2$ will no longer be destructive, since the phase difference between the two rays is no longer $\pi$. As a result, the system will begin to radiate, revealing the presence of the object between the lens and emitter 1. Moreover, a *holographic* type of image of the object will be created since this image results from the interference between the two coherent sources. The properties of this image, however, are beyond the scope of the present study.

The picture based on the concepts of the ray optics, presented so far, is sufficient to reveal the qualitative features of the device. However, a rigorous – full-wave – analysis is needed if any quantitative characterization is to be achieved. This will enable us to assess what potential applications the system might have. The cylindrical symmetry (Fig. 1(a)) reduces the problem to a two-dimensional one. Therefore, the BOR-FDTD method can be used, where BOR stands for bodies-of-revolution [25]. The electromagnetic field components that are not identically zero are $\left(E_\rho, H_\varphi, E_z\right)$ and hence the electromagnetic field configuration is of *E*-type (TM). The Drude model in its standard form is used for both the relative permittivity $\varepsilon$ and permeability $\mu$ of the LHM disc, i.e.

$$\varepsilon(\omega) = \mu(\omega) = 1 - \frac{\omega_p^2}{\omega(\omega + i\nu)}, \quad (2)$$

where $\omega_p$ is the effective plasma frequency, $\nu$ is the collision frequency and $\omega$ is the frequency of the electromagnetic field. Even though the so called *F*-model ([26] and the references therein) represents the frequency dependence of the effective permeability of the split-ring resonator array, the Drude model in the form (2), is widely used in the computational studies [19-22]. Details on the FDTD-modeling of LHMs can be found in [19, 20, 24].

Special attention is paid to the proper modeling of the sources that couple the electromagnetic energy into the LHM-disc. At relatively low excitation frequencies, which is the case studied here, the finite size of the lens is becomes important. This means that the source have to be sufficiently close to the surface of the LHM so that a significant fraction of the total emitted power can be coupled into the lens. As shown in [22], however, strong electromagnetic fields exist near the air-LHM interface stretching towards the image plane, and so, in the case studied here, they may influence the sources themselves. Therefore, the applicability of the standard [25] soft-source model, which amounts to specifying the current density inside the wire, or the hard-source one, which specifies $H_\varphi$ on the surface of the wire and sets $E_z$ to zero inside the wire, for the antennae becomes questionable. Both of them require knowledge of the current distribution inside the antennae and, while in the absence of the lens the latter is known to be sinusoidal [27], such information is not available when the lens is present.

The "improved" version [28] of the thin-wire model [29-31] is used for the dipole antennae. The latter model requires no information for the current distribution inside the wire, since this distribution is calculated in a self-consistent manner. Besides, it allows the calculation of the antenna input impedance that takes into account the finite wire radius $r_0$, thus producing results in a very good agreement with experimental measurements [31].

Knowing the antenna input impedance allows the power balance of the systems to be studied in full detail giving the possibility of evaluating its performance. To appreciate this, consider the Poynting's theorem [27]

$$\mathrm{div}(\boldsymbol{E}\times\boldsymbol{H})+\frac{1}{2}\frac{\partial}{\partial t}\left(\mu_0 H^2+\varepsilon_0 E^2\right)=-\left(\boldsymbol{K}.\boldsymbol{H}+\boldsymbol{J}.\boldsymbol{E}\right)-\boldsymbol{J}_S.\boldsymbol{E}\,, \qquad (3)$$

where $\boldsymbol{J}$ and $\boldsymbol{K}$ are the electric and magnetic current densities associated with the LHM disc and $\boldsymbol{J}_S$ is the current density flowing in the feeding gap of each of the antennae. Considering the time-harmonic feeding voltages, time averaging, and integrating over the spatial variables reduces (3) to

$$P_{in} = P_l + P_r. \tag{4}$$

In (4)

$$P_r = \oint_\Sigma \langle \mathbf{E} \times \mathbf{H} \rangle . d\mathbf{\Sigma} \tag{5}$$

is the total emitted power by the system, in which $\langle . \rangle$ stands for time-averaging and $d\mathbf{\Sigma}$ is a surface element. The power-loss is given by

$$P_l = \int \langle \mathbf{K}.\mathbf{H} + \mathbf{J}.\mathbf{E} \rangle dV \tag{6}$$

and the input power, supplied by the generator to the electromagnetic field is

$$P_{in} = -\int \langle \mathbf{J}_S.\mathbf{E} \rangle dV . \tag{7}$$

Through the input impedance

$$Z(\omega) = \frac{\tilde{U}(\omega)}{\tilde{I}(\omega)}, \tag{8}$$

where $\tilde{U}(\omega)$ and $\tilde{I}(\omega)$ are the Fourier-transforms of the input voltage and the input current, the input power (per dipole) has the form

$$P_{in} = \frac{U_0^2}{2} \frac{R(\omega)}{R^2(\omega) + X^2(\omega)}. \tag{9}$$

In (9), $R(\omega)$ and $X(\omega)$ are input resistance and input reactance, so for a single dipole,

$$Z(\omega) = R(\omega) + iX(\omega). \tag{10}$$

In the presence of losses, which is the case considered here, the input resistance $R$ has two components

$$R = R_l + R_r, \tag{11}$$

with $R_l = R(P_l/P_{in})$ and $R_r = R(P_r/P_{in})$ being the loss resistance and the radiation resistance, respectively.

The computational grid used for the FDTD method deployed here is terminated by a perfectly matched layer (PML) [25]. The integration in (5) is performed over a cylindrical surface, situated sufficiently far from the system, close to the PML region [14]. The resolution $\Delta\rho = \Delta z = \lambda/80$, where $\lambda$ is the excitation wavelength, is used for all the simulations, and the time step $\Delta t$ is set to $\Delta t = \Delta z/4c$, with $c$ being the speed of light in a vacuum. In the absence of a lens the calculations have been benchmarked against [28] with excellent agreement being achieved.

## 3. Numerical results

Fig. 2 shows the input power, the emitted power and the losses, given by Eqs. (7), (5) and (6), respectively, as a function of the lens radius for the system under investigation. Two different values of the collision frequency $\nu$ are used. As Fig. 2(a) and (c) show, the emitted power decreases with the increase of the lens radius, while the losses, associated with dissipation in the LHM disc increase. Fig 2 (c) shows that the sum of the loss power and the radiated power is in good agreement with the results for the input power, which indicates the high level of accuracy of the simulation. The corresponding input, loss and radiation resistances, as well as the input reactance are shown on Fig. 2 (b) and (d). Note, that the radiation resistance ranges from approximately 75 Ω for the zero lens radius case, which is a standard [27] value for a half-wavelength dipole to almost zero at values of the lens radius larger than four wavelengths, indicating that the system becomes non-radiating. At the same time the loss resistance increases from zero and reaches a constant value, which means that almost all of the input power is absorbed in the LHM disc and is not radiated. The radiation resistance is lower at the lower value of the collision frequency, for the same radius, which can be attributed to a stronger electromagnetic field confinement resulting from the better imaging properties of the lens at lower values of $\nu$.

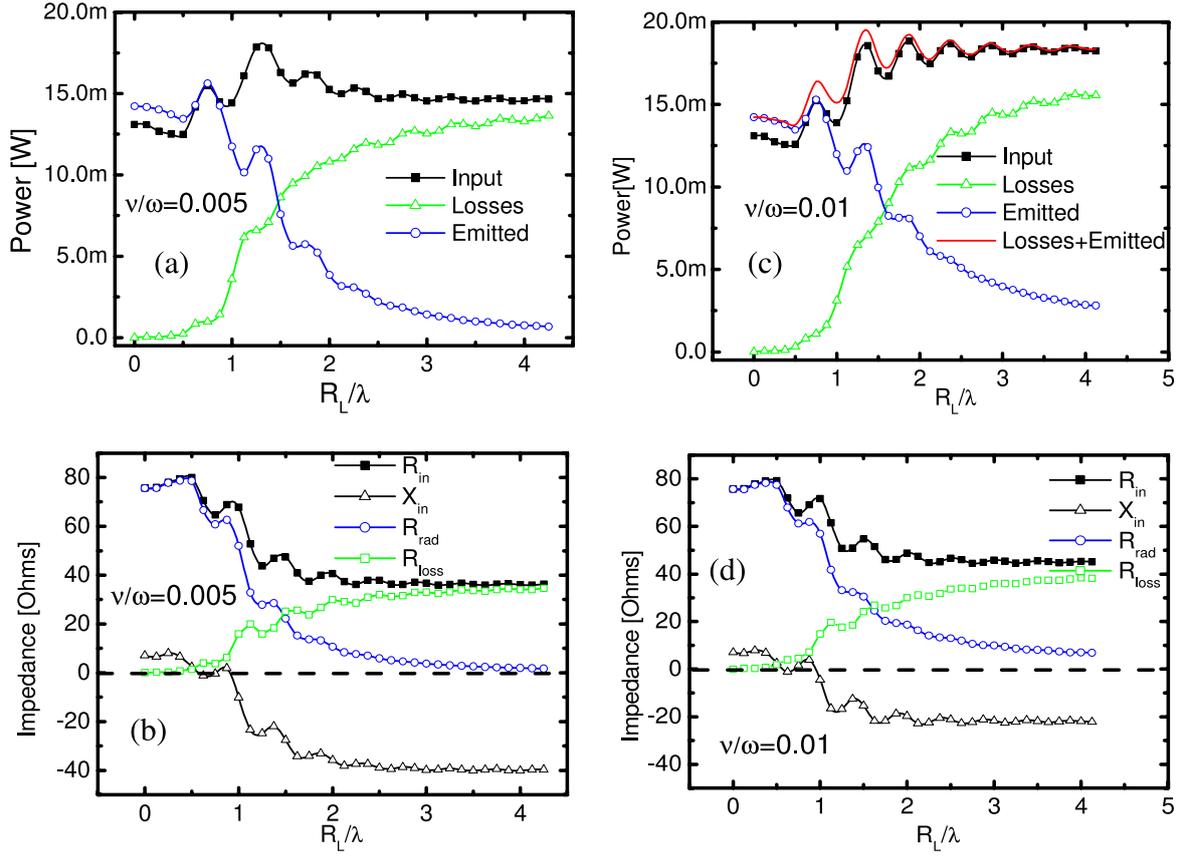

Fig. 2. (Color online) Input power, emitted power and losses as a function of the lens radius for the system shown on Fig. 1 with $\varphi = \pi$ and $U_0 = 1V$ on (a) and (c) at $\nu/\omega = 0.005$ and $\nu/\omega = 0.01$, respectively. The excitation frequency is $\omega/2\pi = 1 GHz$, the dipole length is $L_d = 13.46 cm$ and its radius is $r_0 = L_d/150$. The thickness of the LHM disc is $L = 44.85\ cm$ and the plasma frequency is $\omega_p = \sqrt{2}\omega$. The corresponding dependencies of the input resistance, the radiation resistance, the loss resistance and the input reactance are shown on (b) and (d).

Note that the input reactance of the system increases its absolute value with an increase in the lens radius, which corresponds to decreasing the amount of radiated power. This is in qualitative agreement with earlier results [7, 10] where by means of source decomposition to a sum of a "purely radiating" and non-radiating parts it has been shown that the reactive power of a source is associated mainly with its non-radiating component. On a qualitative level, adding the lens between the two dipoles can be seen as "removing" their radiating part, since in fact no radiation is emitted in the presence of the lens. Adding the lens, however, means that the "source" now consists of two parts: the currents flowing in the antennae and the currents flowing inside the lens. The latter are represented by the current densities $\boldsymbol{K}$ and $\boldsymbol{J}$ in Eq. (3). It has been verified here that the increase of the input reactance corresponds to an increase of the reactive power of the system. Note, that at lower values of the collision frequency, where the properties of the system are closer to that of an "ideal" non-radiating configuration, the absolute value of the input reactance is larger.

Interestingly, all the quantities presented on Fig. 2 display a sequence of local minima and maxima at relatively low values of the lens radius. As can be seen from Fig. 2 two adjacent local maxima (or minima) correspond to an increase of the lens radius with one-half of the excitation wavelength. This behavior is somewhat similar to a "cavity effect" that has been noted previously [22] and occurs when the dimensions of the lens are not much bigger than the wavelength. The cavity effect [22] influences the field distribution inside and on the surface of the lens. Fig. 2 shows that the input resistance and reactance of the emitters *also* become affected. Examining the data shown in Fig. 2 reveals that the maxima occur when the distance between the center of the emitter and the rim of the LHM disc, $D_r = \sqrt{(L/2)^2 + R_L^2}$, is equal to one of the numbers in the sequence $\lambda$, $1.5\lambda$, $2\lambda$, $2.5\lambda$ … . On the other hand local minima occur if $D_r$ takes a value from the set $1.25\lambda$, $1.75\lambda$, $2.25\lambda$, $2.75\lambda$ ..., where $L = 1.5\lambda$. This behavior suggests that standing waves can be formed between the emitter and the rim of the disc. The LHM disc is impedance-matched to the air and no reflections of *radiation* should be possible, however, the radiation is coupled to the "plasmons" [22] that exist on the surface of the lens. These formations can be reflected from the rim of disc, which results in reflecting the radiation coupled to them.

In fact, Fig 2 shows that the input power, which would have been emitted in the absence of the lens, is absorbed in the relatively small volume presented by the lens.

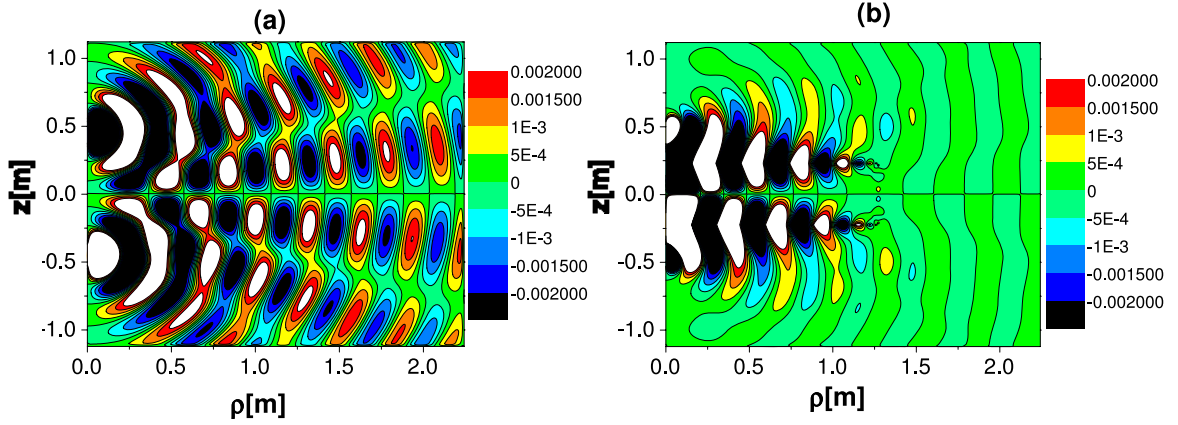

Fig. 3. (Color online) Distribution of the azimuthal magnetic field component $H_\varphi(\rho, z)$ in the absence of the LHM disc ($R_L = 0$, $P_{in} = P_r = 14.22\, mW$) (a) and in its presence ($R_L = 1.27\, m$, $\nu/\omega = 0.005$, $P_{in} = 14.6\, mW$, $P_r = 0.66\, mW$) (b). The black/white color means large negative/positive values of the magnetic field.

This means that strong electromagnetic fields are created in the region between the lens and the two emitters. To verify this Fig. 3 contrasts the magnetic field distribution around the two emitters in the absence of the lens (Fig. 3 (a)) to that in its presence (Fig. 3(b)). The input powers are roughly the same in the two cases but strong electromagnetic fields, shown in black or white colors on Fig. 3, exist only in the region between the two emitters and the lens and inside the lens itself. Outside that region the electromagnetic field is weak. The ability of the system to create strong electromagnetic fields in a limited region of space may be useful for some applications, especially in the medical field.

In order to get an idea about how the system behaves with respect to the inevitable deviations from the "ideal" conditions, Fig. 4 shows the emitted power, the losses and the input power as a function of the excitation frequency for the two different values of collision frequency considered. All the other parameters, including the magnitude of the feeding voltages, are fixed. The emitted power displays a well-defined local minimum around the "resonant" frequency $\omega_0 = 1 GHz$ at which the value of the refractive index of the LHM disc is equal to minus one. This means that the desired property of the system, that it should not radiate electromagnetic energy, does not disappear as a result of an arbitrary small deviation of the index of refraction of the LHM disc from minus one.

The input power is also plotted in the absence of the lens. The comparison of the results in the absence of the lens with those in its presence show that in the frequency range considered the resonant properties [27-30] of the antennae are unimportant. The frequency properties of the system result from the strong frequency dependence of the *coupling* between the two antennas ensured by the LHM. This feature may be useful if the system is used as a frequency etalon.

Fig 4(c) compares the emitted power, normalized with respect to the value at the resonant frequency $\omega_0$ by the system at two different values of the collision frequency $\nu$.

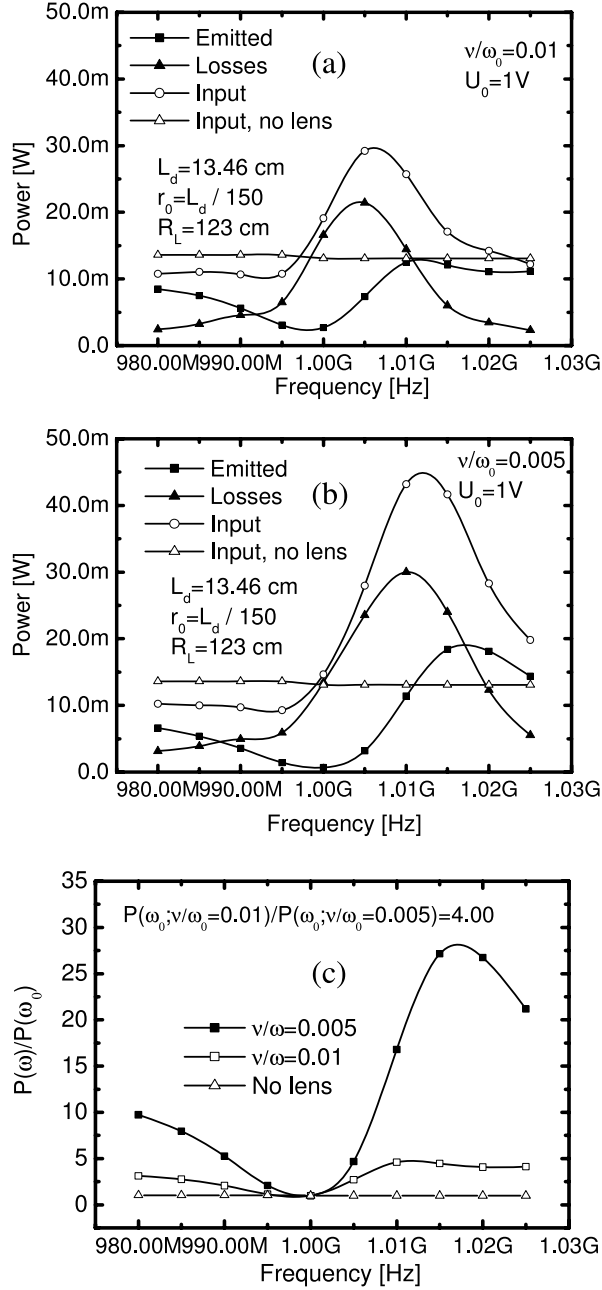

Fig. 4. Input power, radiated power and loss power as a function of the excitation frequency $\omega$. The frequency value at which the index of refraction of the LHM is equal to -1 is denoted by $\omega_0$. The value of the collision frequency is $\nu/\omega_0 = 0.01$ (a) and $\nu/\omega_0 = 0.005$ (b). The emitted power normalized to its value at $\omega_0$ is shown on (c). For a comparison the behavior of the same pair of dipoles but in the absence of the lens is also shown.

The same quantity is also plotted in the absence of the lens. Note that the *relative* changes of the emitted power, for a fixed deviation from the resonant frequency strongly depend on the collision frequency. The lower the collision frequency, the stronger the resonance, as expected physically.

We now turn our attention to the effect of introducing an object of relative dielectric permittivity $\varepsilon_P > 1$ on the radiation properties of the system. Due to the cylindrical symmetry of the model, bodies of revolution are considered only. Nevertheless, generic conclusions can still be drawn.

A ring of internal radius $R_P$, at a distance $Z_P$ from the surface of the lens and with a square-shaped cross section (in the $\rho$-$z$ plane) (Fig. 1) is considered. The power $P_r(R_P, Z_P)$ emitted by the system is calculated as a function of the internal ring radius $R_P$ and the sensitivity of the system is defined as

$$\frac{|\Delta P_r|}{P_r} = \frac{|P_r(R_P, Z_P) - P_r(R_L, Z_P)|}{P_r(R_L, Z_P)}. \tag{12}$$

It is verified that, for the set of parameter values chosen, when the ring is outside the lens ($R_P \geq R_L$) its effect is very weak and the value $P_r(R_L, Z_P)$ is used as a reference value. For comparison, the sensitivity is calculated also in the absence of the lens. The cross section of the ring is an object of "*sub-wavelength*" dimensions and Figure 5 shows some numerical conclusions, where it can be seen that the sensitivity of a *non-radiating* configuration is significantly greater than that of the radiating system that is in place in the absence of the lens. The results shown in Fig. 5 (b) relate to a situation where the center of the ring is closer to the center of emitter 1 (compared to Fig. 5 (a)) and the value of the collision frequency, associated with the losses in the LHM disc, is higher. This results in a lower "quality" of the system. Note, that even when the ring is well within the near zone of emitter 1 (Fig. 1) – the height of the ring is slightly larger than the antenna length – in the absence of the lens this results in about a 0.1 relative change of the emitted power compared to the case when no ring is present. In a similar situation, a value of 2 is obtained for the sensitivity if a non-radiating configuration is used. The sequence of local minima and maxima that can be seen in Fig. 2(b), originate from the standing waves formed between emitter 1 and the ring.

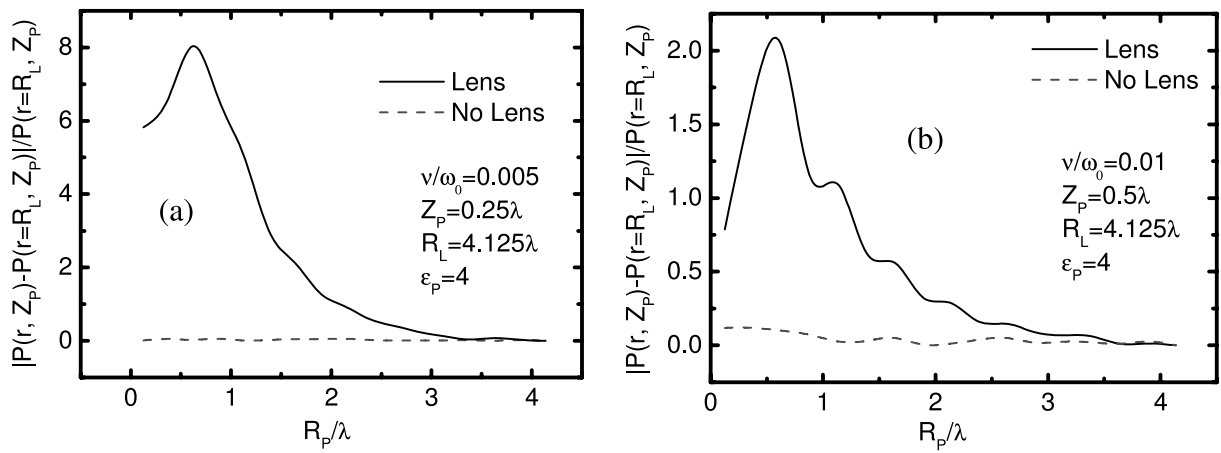

Fig. 5. Comparison between the relative changes in the total emitted power for a non-radiating configuration ("lens") and a pair of dipoles with no lens ("no lens") resulting from the presence of a ring (see Fig. 1) of dielectric permittivity $\varepsilon_P = 4$ and $D = \lambda/2$, where $\lambda$ is the excitation wavelength, as a function of the internal radius of the ring. The frequency is $\omega = 1\,\text{GHz}$. The values of the other parameters are shown on the graphs.

If the center of the ring is farther from the centers of any of the emitters, such as in the case of Fig. 2 (a), then in the absence of the lens, the maximum sensitivity is of the order of 0.02-0.03. In the presence of the lens the maximum sensitivity is 8 (the collision frequency used to generate Fig. 2(a) is lower).

It is clear that an object placed in a region of strong electromagnetic field will change the configuration of that field. However, the important difference between perturbing a propagating field (e.g. the field created by the two antennae in the absence of the lens) and a non-propagating field (the field residing inside a non-radiating configuration) is that in the latter case the object becomes a source of electromagnetic radiation itself. Indeed, there is no radiation in the absence of the object. This radiation can be detected and, in principle, the presence of the object can be revealed. This suggests a possible use of a non-radiating configuration for sensor/detector applications. On the other hand, if the dimensions of the object are smaller than the wavelength, the disturbance (or the "shadow") that will be created by the object, in the case of a propagating field is likely to disappear over a relatively short propagation distance because the diffraction length associated with that disturbance will be small. Generally speaking, it should be easier to distinguish "presence" from "absence" of radiation, especially if large relative changes of the emitted power are associated with the two events (which is the case of a non-radiating configuration) rather than detecting tiny changes on a strong background, which is the case if propagating fields are transmitted through the object. Practical situations, where the latter feature will be useful may involve detecting the presence of small (sub-wavelength) particles in the ambient material. Another area of possible use may include some special security applications. By definition a non-radiating configuration does not create electromagnetic fields outside its own volume, which makes it hard to detect from outside. The only possibility of detecting it is by disturbing the non-propagating field inside, but as we have shown, this will result in radiation emitted by the system, which will trigger the appropriate event.

The non-radiating system described so far is based on the destructive interference between the electromagnetic fields created by two identical wire dipoles. This is achieved by using a disc of LHM to couple the radiation produced by the two emitters, which are driven by $\pi$-out-of-phase voltages. In what follows the case of in-phase driving voltages is considered and the interference between the fields created by the two dipoles is constructive in this case. The question that is addressed is whether the coupling effect can improve the *individual* radiation properties of each of the dipoles.

Figure 6 shows the input impedance together with the radiation resistance and the loss resistance as functions of the antenna length in three cases at a fixed value of the excitation frequency: The cases considered include (i) two dipoles coupled by the LHM disc, (ii) one of the dipoles being switched off so that only one of them is radiating and the LHM disc is present (iii) a single dipole emitting in the absence of the lens. Note, that apart from the range of very short antennae, introducing the LHM disc reduces the absolute value of the input reactance. Figure 6(a) allows three different regimes to be identified. When the antenna length is in the range between 1 and 1.4 wavelengths the behavior of a single dipole, emitting near the LHM disc and that of two dipoles acting together is similar. At the same time, the values of the input impedance in this regime are different from those obtained in the absence of the LHM disc. This means that each of the *individual* dipoles is affected by the presence of the LHM but the coupling between the two antennae is negligible. For antenna lengths in the range between 0.6 and 1 of a wavelength, all the three systems behave in a different way from each other, since different values are obtained for the input impedance. However, if a single dipole is emitting the *position* of the resonance (maximum input resistance and zero reactance) remains unaffected by the presence of the LHM. The resonant values of the input resistance are indeed different in the two cases (with or without a LHM disc) but the resonance occurs at 0.8 of a wavelength, regardless of whether a LHM disc is present or not. The losses in the disc decrease the strength of the resonance, but have little effect on the resonant frequency. The third regime, the regime of relatively short antennas (antenna length less than 0.6 wavelengths) is the most interesting one, and the results obtained in this case are presented in somewhat more detail on

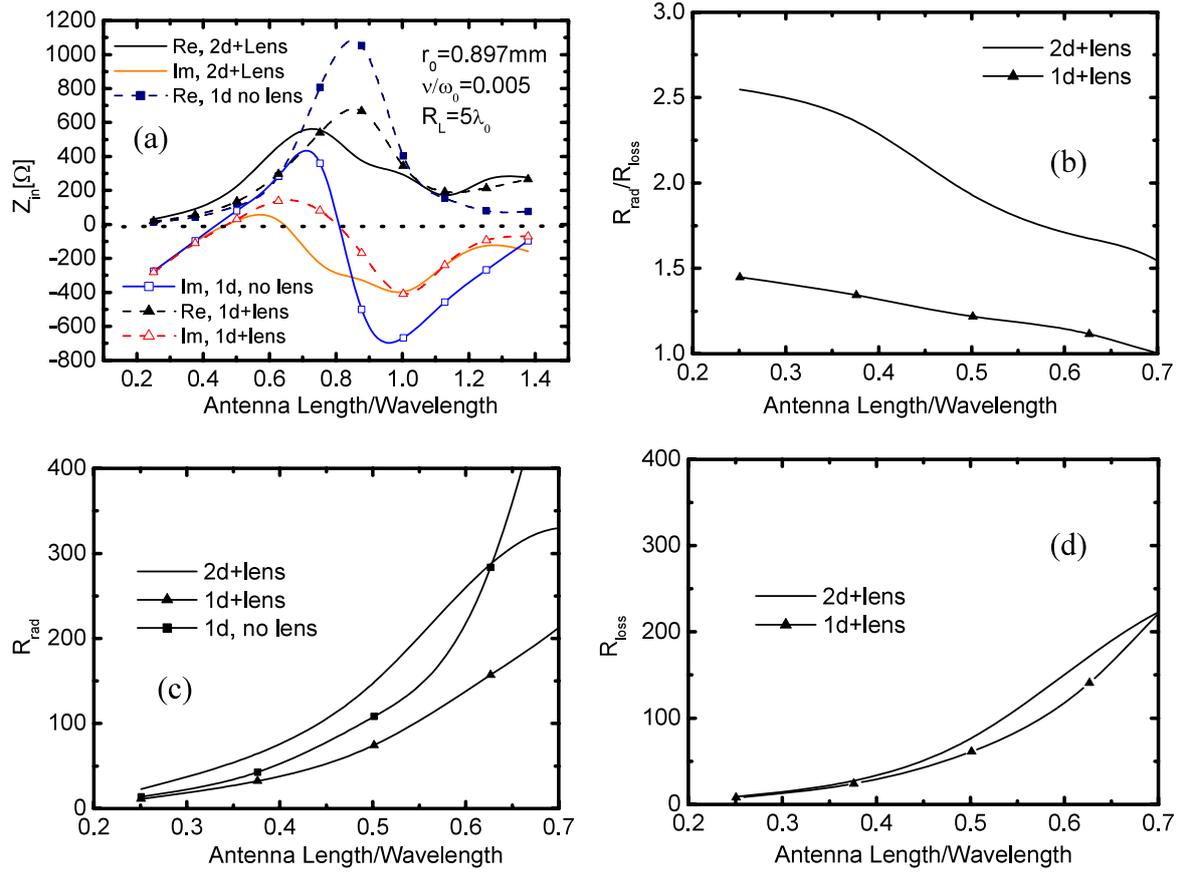

Fig. 6. (Color online) Input impedance of each of the dipoles in the system of Fig. 1 but driven with in-phase voltages as a function of the antenna length. The excitation frequency is fixed to 1 GHz, the value at which the index of refraction of the LHM disc is equal to minus 1. The input resistance ("Re") and input reactance ("Im") are shown in the following three cases on (a): both the dipoles are emitting and the lens is present ("2d+lens"); single dipole emitting in the presence of the lens ("1d+lens"); single dipole emitting in the absence of the lens ("no lens"). The ratio between the radiation resistance and the loss resistance in the presence of the lens for two dipoles emitting simultaneously ("2d+lens") and for a single dipole ("1d+lens") on (b). Radiation resistance in the presence of the lens with two dipoles radiating ("2d+lens") and one dipole radiating ("1d+lens") and of a single dipole in the absence of the lens ("1d, no lens") on (c). Loss resistance in the presence of the lens with single dipole radiating ("1d+lens") and the two dipole radiating simultaneously ("2d+lens") on (d).

Fig. 6 (b)-(d). First of all if the two antennas coupled by the lens are emitting together the resonance is shifted towards lower antenna lengths (Fig. 6 (a)). This results in an increase of the *individual* input resistance of each of the two dipoles, compared to the case when a single dipole is acting alone. In contrast, if a single dipole is emitting, its input resistance obtained in the presence of the LHM and in its absence is practically the same value. Therefore coupling two relatively *short* dipole antennae leads to an improvement of their individual characteristics. In order to verify this conclusion the two components of the input resistance, namely the loss resistance and the radiation resistance, are considered separately in the regime of short antenna lengths. As shown in Fig 6 (b) the ratio between the radiation resistance and the loss resistance is larger if two dipoles are used. Besides that, Fig. 6 (c) shows that with one of the dipoles switched off the radiation resistance of the remaining one is even lower than that of a dipole with no LHM present. This indicates that the role of the LHM with only one *short* dipole emitting is reduced to introducing losses in the system.

Figure 6 (d) shows that the loss resistance is larger in the case when two dipoles are emitting together than that of a single dipole. The increase of the loss resistance is, however, accompanied by a *stronger* increase of the radiation resistance, as Fig. 6(b) shows.

It should be emphasized that the use of the thin-wire model [28-31] for the purposes of the present study is important. The use of the simpler "hard-" or "soft-source" models [25] requires *a priori* information about the current distribution in the wire. This is quite different from the case of a single dipole and no LHM disc because then the current distribution is known [27]. No such information is available in the presence of the LHM disc.

It has been shown recently [32, 33] that if an infinitesimal electric dipole is placed in a suitably chosen spherical shell made of an LHM, its radiation properties can be improved significantly. Here a similar effect is reported. In this case, however, the improvement originates from the coupling between *two* antennae, rather than from the presence of the LHM disc itself. As can be seen from Fig. 6 (a) the radiation properties of small antennae are poor – small radiation resistance and large reactance are obtained in this case. An attempt to seek an improvement is driven by demands for miniaturization.

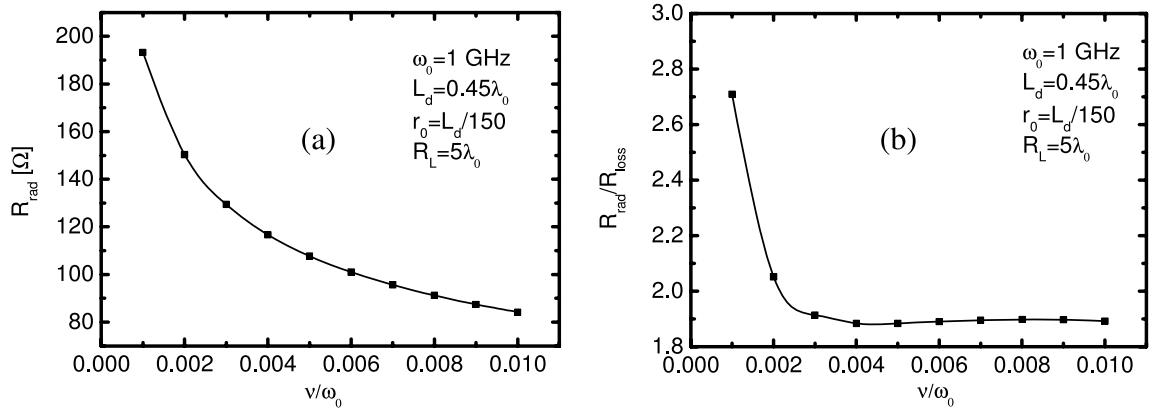

Fig. 7. Radiation and loss resistance for a system of wire dipoles coupled through a disc of LHM and fed with in-phase voltages. In the absence of any lens the radiation resistance of the dipole is 75 $\Omega$. Radiation resistance as a function of the collision frequency in (a). The ratio between the radiation and the loss resistances as a function of the collision frequency in (b).

The effect of the losses on the performance of the system of the two dipoles coupled by the lens is shown on Figure 7. As Fig. 7(a) shows, the radiation resistance increases with the *decrease* of the collision frequency. At the same time, as seen in Fig. 7(b), the ratio between the radiation resistance and the loss resistance remains unaffected in a range of collision frequency values that map onto a fivefold increase of the collision frequency. The ratio between the radiation resistance increases at lower values of the collision frequency. This increase can be understood by acknowledging that in the absence of losses the loss resistance must be zero. However, the case of very low collision frequency values is difficult to analyze numerically, because the system takes very long time to reach a stationary state [21].

## 4. Conclusions

It is shown that the remarkable properties of left-handed metamaterials can be used to create a non-radiating electromagnetic configuration. The properties of the system proposed are studied in full numerical detail with the aim of identifying what possible applications such a system might have. It is shown that if the non-propagating electromagnetic field is disturbed by the intrusion of any object into the system it will begin to radiate, thus revealing the presence of that object. This combined with the property that a non-radiating configuration remains unaffected by objects outside the region occupied by its own non-propagating field suggests far-reaching possibilities. It is shown that coupling two identical short wire-dipoles, fed with in-phase voltages, with a disc made of LHM improves their radiation properties.

**Acknowledgements**

This work is supported by the Engineering and Physical Sciences Research Council (UK) under the Adventure Fund Programme.

**References**

1. G. H. Goedecke, "Classically radiationless motions and possible implications for quantum theory" Phys. Rev. **135**, B281-B288 (1964).

2. G. B. Arnett and G. H. Goedecke, "Electromagnetic fields of accelerated nonradiating charge distributions" Phys. Rev. **168**, 1424-1428 (1968).

3. D. Bohm and M. Weinstein, "The self-oscillations of a charged particle", Phys. Rev. **74**, 1789-1798 (1948).

4. A. J. Devaney and E. Wolf, "Radiating and nonradiating classical current distributions and the fields they generate", Phys. Rev. D **8**, 1044-1047 (1973).

5. E. A. Marengo and A. J. Devaney, "Nonradiating sources with connections to the adjoint problem", Phys. Rev. E **70**, 037601 (2004).

6. K. Kim and E. Wolf, "Nonradiating sources and their fields", Opt. Commun. **59**, 1-6 (1986).

7. E. A. Marengo and R. W. Ziolkowski, "On the radiating and nonradiating components of a scalar, electromagnetic and weak gravitational sources", Phys. Rev. Lett. **83**, 3345-3349 (1999).

8. A. J. Devaney and G. C. Sherman, "Nonuniqueness in inverse-source and inverse scattering poblems", IEEE Trans. on Antennas and Propagation **AP-30**, 1034-1037 (1982).

9. E. A. Marengo and R. W. Ziolkowski, "Inverse source problem with regularity constraints: normal solution and nonradiating source components", J. Opt. A: Pure Appl. Opt. **2**, 179-187 (2000).

10. E. A. Marengo and R. W. Ziolkowski, "Nonradiating and minimum energy sources and their fields: generalized source inversion theory and applications", IEEE Trans. on Antennas and Propagation **48**, 1553-1562 (2000).

11. N. K. Nikolova and Y. S. Rickard, "Nonradiating electromagnetic sources in a nonuniform medium", Phys. Rev. E **71**, 016617 (2005).

12. G. N. Afanasiev and V. M. Dubovik, "Some remarkable charge-current configurations", Phys. Part. Nuclei **29**, 366-391 (1998).

13. G. N. Afanasiev and Yu. P. Stepanovsky, "The electromagnetic field of elementary time-dependent toroidal sources", J. Phys. A: Math. Gen. **28**, 4565-4580 (1995).

14. A. D. Boardman, K. Marinov, N. Zheludev and V. A. Fedotov, "Dispersion properties of nonradiating configurations: Finite-difference time-domain modeling" Phys. Rev. E **72,** 036603 (2005).

15. A. D. Boardman, K. Marinov, N. Zheludev and V. A. Fedotov, "Nonradiating toroidal structures", in Metamaterials, T. Szoplik, E. Ozbay, C. M. Soukoulis and N. I. Zheludev, eds., Proc. SPIE **5955**, 595504 (2005).

16. R. A. Silin, *Opt. Spectrosc.* (USSR) **44**, "Possibility to produce plane-parallel lenses" 109-110 (1978).

17. V. G. Veselago, "The electrodynamics of substances with simultaneously negative values of ε and μ", Sov. Phys. Usp. **10**, 509-514, (1968).

18. J. B. Pendry, "Negative refraction makes a perfect lens" *Phys. Rev. Lett.* **85**, 3966-3969 (2000).

19. R. Ziolkowski, "Pulsed and CW Gaussian beam interactions with double negative metamaterial slabs", *Opt. Express* **11**, 662-681 (2003).

20. R. Ziolkowski and E. Heyman, "Wave propagation in media having negative permittivity and permeability", *Phys. Rev. E* **64**, 056625 (2001).

21. X. S. Rao and C. K. Ong, "Subwavelength imaging by a left-handed material superlens", *Phys. Rev. E.* **68**, 067601 (2003).

22. L. Chen, S. He and L. Shen, "Finite-size effects of a left-handed material slab on the image quality", Phys. Rev. Lett. **92**, 107404 (2004).

23. G. V. Eleftheriades and K. G. Balmain, *Negative-Refraction Metamaterials: Fundamental Principles and Apllications* (Wiley, Hoboken, NJ, 2005).

24. A. D. Boardman, N. King and L. Velasco, "Negative refraction in perspective", Electromagnetics **25**, 365-389 (2005).

25. A. Taflove and S. Hagness, *Computational Electrodynamics: The Finite-Difference Time-Domain Method* (Artech House, Norwood, MA 2000).

26. R. Ruppin, "Surface polaritons of a left-handed material slab", J. Phys.: Condens. Matter **13**, 1811-1819 (2001).

27. J. A. Kong, *Electromagnetic Wave Theory* (Wiley, New York, 1990)

28. So-ichi Watanabe, "An improved FDTD model for the feeding gap of a thin-wire model", IEEE Microwave and Guided Wave Lett **8**, 152-154 (1998).

29. K. R. Umashankar, A. Taflove and B. Beker, "Calculation and experimental validation of induced currents on coupled wires in an arbitrary shaped cavity", IEEE Trans. Antennas Propag. **AP-35**, 1248-1257 (1987).

30. A. Taflove, K. R. Umashankar, B. Beker, F. Harfoush and K. S. Yee, "Detailed FD-TD analysis of electromagnetic fields penetrating narrow slots and lapped joints in thick conducting screens" IEEE Trans. Antennas Propag. **36**, 247-257 (1988).

31. R. Luebbers, L. Chen, T. Uno and S. Adachi "FDTD calculation of radiation patterns, impedance and gain for a monopole antenna on a conducting box" IEEE Trans. Antennas Propag. **40**, 1577-1583, (1992).

32. R. Ziolkowski and A. Kipple, "Reciprocity between the effects of resonant scattering and enhancent radiated power by electrically small antennas in the presence of nested metamaterial shells", Phys. Rev. E **72**, 036602 (2005).

33. R. Ziolkowski and A. Kipple, "Application of double negative materials to increase the power radiated by electrically small antennas", IEEE Trans. Antennas Propag. **51**, 2626-2640 (2003).